\documentclass[a4paper,12pt]{article}
\usepackage{amsmath}
\usepackage{amssymb}
\usepackage{amsfonts}
\usepackage{bbm} 
\usepackage{fleqn} 
\usepackage[footnotesize]{caption}
\usepackage{graphicx} 
\usepackage{braket} 
\usepackage{mathrsfs}
\usepackage[center,footnotesize,hang]{subfigure}

\newcommand{\CenterEps}[2][1]{\ensuremath{\vcenter{\hbox{\includegraphics[scale=#1]{#2.eps}}}}}
\newcommand{\SuperField}[1]{\hat{#1}}
 
\def\Ng{j} 
\def\Nf{i}

\def\I{\mathrm{i}}
\def\SU{\text{SU}}
\def\SO{\text{SO}}

\def\<{\left\langle}
\def\>{\right\rangle}
\def\ChargeC{\mathrm{C}} 
\def\chargec{\mathrm{C}}

\DeclareMathOperator{\Tr}{Tr}

\begin{document}

\bibliographystyle{OurBibTeX}

\begin{titlepage}

 \vspace*{-15mm}
\begin{flushright}
SHEP/0415
\end{flushright}
\vspace*{5mm}

\begin{center}
{
\sffamily
\LARGE
Type II Leptogenesis and the Neutrino Mass Scale
}
\\[12mm]
S. Antusch\footnote{E-mail: \texttt{santusch@hep.phys.soton.ac.uk}},
S. F. King\footnote{E-mail: \texttt{sfk@hep.phys.soton.ac.uk}}
\\[1mm]
{\small\it
Department of Physics and Astronomy,
University of Southampton,\\
Southampton, SO17 1BJ, U.K.
}
\end{center}
\vspace*{1.00cm}

\begin{abstract}

\noindent 
We discuss the effect of the neutrino mass scale on  
baryogenesis via the out-of-equilibrium decay 
of the lightest right-handed (s)neutrinos
in type II see-saw models. 
We calculate the type II contributions to the decay asymmetries
for minimal scenarios based on the Standard Model and on the 
Minimal Supersymmetric Standard Model, 
where the additional direct mass term for the neutrinos 
arises from a Higgs triplet vacuum expectation value. 
The result in the supersymmetric case is new 
and we correct the previous result in the scenario based on the 
Standard Model.  
We confirm and generalize our results by calculating the decay asymmetries in 
an effective approach, which is independent of the realization 
of the type II contribution. 
We derive a general upper bound on the 
decay asymmetry in type II see-saw models 
and find that it increases with the neutrino mass scale, 
in sharp contrast to the type I case which leads to an upper bound 
of about 0.1 eV on the neutrino mass scale. 
We find a lower bound on the mass of the lightest right-handed neutrino, 
significantly below the corresponding type I bound for partially 
degenerate neutrinos. This lower bound decreases 
with increasing neutrino mass scale, making
leptogenesis more consistent with the gravitino constraints 
in supersymmetric models.

\end{abstract}

\end{titlepage}
\newpage
\setcounter{footnote}{0}

\section{Introduction}
Leptogenesis \cite{Fukugita:1986hr} is one of the most attractive mechanisms 
for explaining the observed baryon asymmetry of the universe, 
$n_\mathrm{B} /n_\gamma 
= (6.5^{+0.4}_{-0.8}) \cdot 10^{-10}$ \cite{Spergel:2003cb}. 
In the type I see-saw scenario \cite{seesaw},  
where the asymmetry is generated via the 
out-of-equilibrium 
decay of the same heavy right-handed neutrinos which are involved in 
generating neutrino masses, it has been studied intensively.  
In models with a left-right 
 symmetric particle content like minimal left-right 
symmetric models, Pati-Salam models or Grand Unified Theories (GUTs) 
based on $\SO (10)$, the type I see-saw mechanism is typically 
generalized to a type II see-saw (see e.g.\ \cite{seesaw2}), 
where an additional direct 
mass term $m_{\mathrm{LL}}^{\mathrm{II}}$ for the light neutrinos is present. 
The effective mass matrix of the light neutrinos is then given by 
\begin{eqnarray}\label{eq:SeeSawFormula}
m^{\nu}_{\mathrm{LL}} = 
m_{\mathrm{LL}}^{\mathrm{I}}+ m_{\mathrm{LL}}^{\mathrm{II}}\;,\;\;
\mbox{where}\;\;
m_{\mathrm{LL}}^{\mathrm{I}} = - v^2_\mathrm{u}\, 
Y_\nu \, M_\mathrm{RR}^{-1}\, Y_\nu^T 
\end{eqnarray} 
is the type I see-saw mass matrix. One motivation for considering  
the type II see-saw is that it allows to construct models for partially
degenerate neutrinos in a natural way, e.g.\ via a type II upgrade 
\cite{Antusch:2004xd}, which is otherwise difficult to achieve in 
type I models. 
From a rather model independent viewpoint, the type II mass term can 
be considered as an additional contribution to the lowest dimensional 
effective neutrino mass operator. 

In the literature, the most discussed case is where the type II contribution
is realized via SU(2)$_{\mathrm{L}}$-triplets. There are in general two
possibilities to generate the baryon asymmetry: via the decay of the lightest
right-handed neutrino $\nu^1_{\mathrm{R}}$ or via the decay of one or more 
SU(2)$_{\mathrm{L}}$-triplets \cite{O'Donnell:1994am,Ma:1998dx,Hambye:2000ui}. 
In the first case, there are additional one-loop diagrams where 
virtual triplets are running in the loop 
\cite{O'Donnell:1994am,Lazarides:1998iq,Chun:2000dr,Hambye:2003ka}. 
Referring to the 
contributions to the decay asymmetries for $\nu^1_{\mathrm{R}}$ 
proportional to $m_{\mathrm{LL}}^{\mathrm{I}}$ as 
$\varepsilon^\mathrm{I}_1$ and to the ones proportional to 
$m_{\mathrm{LL}}^{\mathrm{II}}$ as $\varepsilon^\mathrm{II}_1$, either or both 
contributions can be important for generating the baryon asymmetry. 
In many studies of leptogenesis in left-right symmetric models, it has been
assumed that $\varepsilon^\mathrm{I}_1$ dominates even if neutrino masses stem
dominantly from $m_{\mathrm{LL}}^{\mathrm{II}}$ (see e.g.\ 
\cite{Joshipura:1999is,Joshipura:2001ya,Joshipura:2001ui,Rodejohann:2002hx}). 
The case where $\varepsilon^\mathrm{II}_1$ dominates over
$\varepsilon^\mathrm{I}_1$ has recently been studied in 
\cite{Hambye:2003ka,Rodejohann:2004cg}. 
It has the interesting feature that unlike in the type I see-saw scenario, 
there is in general no upper bound on the absolute neutrino mass scale 
\cite{Buchmuller:2003gz} from type II leptogenesis, as has been pointed out in 
\cite{Hambye:2003ka}.

In this work, we analyze the consequences of the neutrino mass scale for 
baryogenesis via the out-of-equilibrium decay 
of the lightest right-handed (s)neutrinos
in type II see-saw models.
First, we calculate the type II contributions to the decay asymmetries
for minimal scenarios based on the Standard Model (SM) and on the 
Minimal Supersymmetric Standard Model (MSSM), 
where the additional direct mass term for the neutrinos 
stems from the induced vev of a triplet Higgs. 
The result for the supersymmetric case is new 
and we correct the previous result in the scenario based on the Standard Model.  
We then develop an effective approach to type II leptogenesis, 
assuming a gap between the mass 
$M_{\mathrm{R}1}$ of the lightest
(s)neutrino and the masses of the heavier particles involved in generating 
neutrino masses. 
Leptogenesis in this framework is 
approximately independent of the specific realization of the neutrino mass operator. 
The calculation of the decay asymmetries using 
the effective approach confirms our results for the 
triplet scenarios in the limit of heavy triplets.  
We subsequently derive a general upper bound on the 
decay asymmetry and find that it increases with the neutrino mass scale. 
It leads to a lower bound on the mass of the lightest right-handed neutrino, 
which is significantly below the type I bound for partially 
degenerate neutrinos.  
It is worth emphasizing that these results are in sharp contrast to the type I 
see-saw mechanism where an upper bound on the neutrino mass scale is predicted. 
Here we find no upper limit on the neutrino mass scale which may be increased
arbitrarily. Indeed we find that the lower bound on the mass of the lightest
right-handed neutrino decreases as the physical neutrino mass scale increases.  
This allows a lower reheat temperature, making thermal leptogenesis more consistent 
with the gravitino constraints in supersymmetric models 
\cite{Khlopov:1984pf,Ellis:1984eq,Ellis:1985er,Moroi:1993mb}).

\section{Decay Asymmetries for Type II via Triplets}\label{sec:triplets}\label{sec:TripletDecayAsymmetries}
We now consider minimal type II models based on
the SM and the MSSM, 
where the type II see-saw is realized via an additional heavy 
SU(2)$_{\mathrm{L}}$-triplet. 
We focus on the case where the asymmetry is generated 
via the decay of the lightest right-handed (s)neutrinos and   
assume hierarchical masses of the right-handed (s)neutrinos and 
$M_\Delta \gg M_{\mathrm{R}1}$.  

\subsection{Minimal Type II See-Saw Scenarios}
In the MSSM extended by chiral superfields 
$\SuperField{\nu}^{\ChargeC \Nf}$ $(\Nf\in \{1,2,3\})$, 
which contain the right-handed neutrinos $\nu^{\Nf}_\mathrm{R}$ and 
 $\SU (2)_\mathrm{L}$-triplet Higgs superfields 
$\SuperField{\Delta}$ 
and $\SuperField{\bar\Delta}$ with weak hypercharge $1$ and $-1$, respectively, 
the relevant parts of the superpotential are 
\begin{subequations}\begin{eqnarray}
 \mathcal{W}^{\mathrm{MSSM}}_{\nu^{\chargec}} & = &
(Y_\nu)_{f \Ng}(\SuperField{L}^{f}\cdot
 \SuperField{H}_\mathrm{u})\,  \SuperField{\nu}^{\chargec \Ng} 
 + \frac{1}{2} \SuperField{\nu}^{\chargec \Nf} (M_\mathrm{RR})_{\Nf \Ng}
 \SuperField{\nu}^{\chargec \Ng}
 \; , \\
 \mathcal{W}^{\mathrm{MSSM}}_{Y_\Delta} & = &
 \frac{1}{2} \,(Y_\Delta)_{fg} \,\SuperField{L}^T{}^f\,i\tau_2\, \SuperField{\Delta} 
 \,\SuperField{L}^g\; , \\
 \mathcal{W}^{\mathrm{MSSM}}_{\Delta,H} & = & 
 M_\Delta \Tr (\SuperField{\Delta} \,\SuperField{\bar\Delta})
+ \lambda_\mathrm{u}  \,
\SuperField{H}^T_\mathrm{u}\,i\tau_2\, \SuperField{\bar\Delta}\,\SuperField{H}_\mathrm{u}
 +\lambda_\mathrm{d}\,
\SuperField{H}^T_\mathrm{d}\,i\tau_2\, \SuperField{\Delta}\,
\SuperField{H}_\mathrm{d} \; .
\end{eqnarray}\end{subequations}
The dot indicates the $\SU (2)_\mathrm{L}$-invariant product, 
$(\SuperField{L}^{f}\cdot \SuperField{H}_\mathrm{u}) := 
\SuperField{L}_a^{f}(i\tau_2)^{ab} (\SuperField{H}_\mathrm{u})_b$, with $\tau_A$  
$(A\in \{1,2,3\})$ being
the Pauli matrices.   
Superfields are marked by hats and we have written the 
$\mathrm{SU}(2)_\mathrm{L}$-triplets 
as traceless $2\times2$-matrices 
\begin{eqnarray}
 \SuperField{\Delta}= 
\left(\begin{array}{cc}
 \SuperField{\Delta}^{+}/\sqrt{2}& \SuperField{\Delta}^{++} \\
 \SuperField{\Delta}^{0}  &-\SuperField{\Delta}^{+}/\sqrt{2} 
	 \end{array}\right)  \:\mbox{and}\;\;
\SuperField{\bar\Delta}= 
 \left(\begin{array}{cc}
 \SuperField{\bar\Delta}^{+}/\sqrt{2}& \SuperField{\bar\Delta}^{++} \\
 \SuperField{\bar\Delta}^{0}  &-\SuperField{\bar\Delta}^{+}/\sqrt{2} 
	 \end{array}\right)  .	 
\end{eqnarray}
In the SM, we only consider one triplet scalar field $\Delta$, 
using an analogous notation as
in the extended MSSM.  
The corresponding terms of the Lagrangian are
\begin{subequations}\begin{eqnarray}
 \mathcal{L}^{\mathrm{SM}}_{\nu_\mathrm{\mathrm{R}}} & = &
- (Y_\nu)_{f \Ng}({L}^{f}\cdot
 H)\,  \nu_\mathrm{\mathrm{R}}^{\Ng} 
 - \frac{1}{2} \overline{\nu_{\mathrm{R}}^{\Nf}} (M_\mathrm{RR})_{\Nf \Ng}
 \nu_\mathrm{R}^{\ChargeC\Ng}\;+\;\text{h.c.}
 \; , \\
 \mathcal{L}^{\mathrm{SM}}_{Y_\Delta} & = &
 - \frac{1}{2} \,(Y_\Delta)_{fg} \,{L}^T{}^f\,i\tau_2\, {\Delta} 
 \,{L}^g\;+\;\text{h.c.}\;, \\
 \mathcal{L}^{\mathrm{SM}}_{\Delta,H} & = & 
 - M^2_\Delta \Tr (\Delta^\dagger \Delta)
- \Lambda_\mathrm{u}  \,
H^T\,i\tau_2\, \Delta^\dagger\,H
 \;+\;\text{h.c.}\; . \vphantom{\frac{1}{2}}
\end{eqnarray}\end{subequations}
At low energy in the SM and in the MSSM,  
the type I contribution to the neutrino mass matrix of the light neutrinos 
is approximately given by the see-saw formula of equation 
(\ref{eq:SeeSawFormula}),   
\begin{eqnarray}\label{eq:TypISeeSawFormula}
m^{\mathrm{I}}_\mathrm{LL} = - \,  v^2_\mathrm{u}\, 
Y_\nu \, M_\mathrm{RR}^{-1}\, Y_\nu^T \; .
\end{eqnarray} 
$v_\mathrm{u}$ is the vacuum expectation value (vev) of the neutral component of
the Higgs doublet 
which couples to the right-handed neutrinos and the lepton doublets,
i.e.~$v_\mathrm{u} = \<H^0_\mathrm{u}\>$ in the MSSM and 
$v_\mathrm{u} = \< H^0 \>$ in the SM. 
An induced see-saw suppressed vev $v_\Delta=\<{\Delta}^0\>$ of the neutral 
component of the scalar field contained 
in $\SuperField{\Delta}$ gives a naturally small direct mass for the 
left-handed neutrinos. 
It can also be viewed as resulting from realizing the effective neutrino
mass operator by integrating out the triplet below its mass threshold at 
$M_\Delta$. 
The type II contribution to the effective neutrino mass matrix 
is given by
\begin{eqnarray}
m^{\mathrm{II}}_{\mathrm{LL}} = (Y_\Delta v_\Delta)^* \; , \;\;
\mbox{with}\;\;
v^{\mathrm{SM}}_\Delta:= v_\mathrm{u}^2 \Lambda_\mathrm{u} M^{-2}_\Delta 
\; \; \mbox{and} \; \;
v^{\mathrm{MSSM}}_\Delta:= v_\mathrm{u}^2 \lambda_\mathrm{u} M^{-1}_\Delta\;.
\end{eqnarray} 
 The complete neutrino mass matrix in the minimal type II scenarios based 
on the SM and on the MSSM is thus 
given from the above equations as 
\begin{eqnarray}
m^{\nu}_{\mathrm{LL}} = 
m^{\mathrm{I}}_{\mathrm{LL}} + m^{\mathrm{II}}_{\mathrm{LL}} 
= - v^2_\mathrm{u}\, 
Y_\nu \, M_\mathrm{RR}^{-1}\, Y_\nu^T + (Y_\Delta v_\Delta)^*\; .
\end{eqnarray}

\subsection{Results for the Decay Asymmetries}
 
\begin{figure}
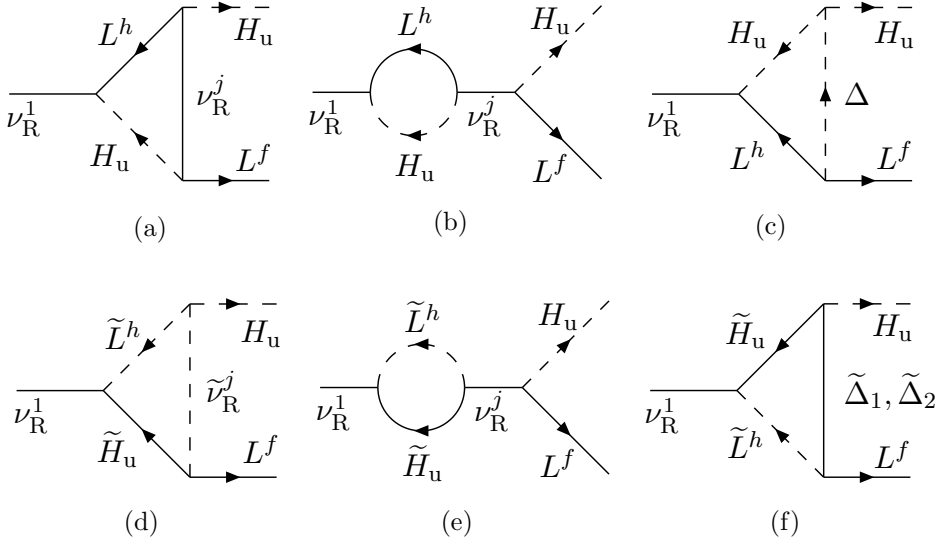

 \begin{center}  
 \CenterEps[1]{TypeIILoops}
 \end{center}
 \caption{\label{fig:Leptogenesis_Triplet_SUSY}
 Loop diagrams in the MSSM which contribute to the decay 
 $\nu^1_{\mathrm{R}}\rightarrow L^f_a H_\mathrm{u}{}_b$ for the case of a 
 type II see-saw mechanism 
 where the direct mass term for the neutrinos stems from the induced vev of a 
 Higgs triplet. 
 In diagram (f), 
 $\widetilde \Delta_1$ and $\widetilde \Delta_2$ are the mass eigenstates
 corresponding to the superpartners of the 
 SU(2)$_{\mathrm{L}}$-triplet scalar fields $\Delta$ and $\bar{\Delta}$.
 The SM diagrams are the ones where no superpartners (marked  
 by a tilde) are involved and where $H_\mathrm{u}{}$ is renamed to the SM Higgs. 
 }
\end{figure}

In this subsection we calculate the relevant decay asymmetries 
diagrammatically. 
The asymmetry from the decay of the lightest right-handed neutrino 
into a lepton doublet and a Higgs is defined as 
\begin{eqnarray}
\varepsilon_{1} := 
\frac{
\Gamma_{\nu^1_{\mathrm{R}} L} - \Gamma_{\nu^1_{\mathrm{R}} \overline L}
}{
\Gamma_{\nu^1_{\mathrm{R}} L} + \Gamma_{\nu^1_{\mathrm{R}} \overline L}
}\; ,
\end{eqnarray} 
with the decay rate $\Gamma_{\nu^1_{\mathrm{R}} L}:= 
\sum_{a,b}\Gamma (\nu^1_{\mathrm{R}}\rightarrow L^f_a H_\mathrm{u}{}_b)$. 
In addition, in the supersymmetric case, we need the decay asymmetries
\begin{eqnarray}
\widetilde  \varepsilon_{1} :=
\frac{
\Gamma_{\nu^1_{\mathrm{R}} \widetilde L} - \Gamma_{\nu^1_{\mathrm{R}}
\widetilde{L}^*}
}{
\Gamma_{\nu^1_{\mathrm{R}} \widetilde L} + \Gamma_{\nu^1_{\mathrm{R}} 
\widetilde{L}^*}
}\; , \quad
\varepsilon_{\widetilde 1} := 
\frac{
\Gamma_{\widetilde \nu^{1*}_{\mathrm{R}} L} - \Gamma_{\widetilde \nu^1_{\mathrm{R}} \overline L}
}{
\Gamma_{\widetilde \nu^{1*}_{\mathrm{R}} L} + \Gamma_{\widetilde \nu^1_{\mathrm{R}} \overline L}
}\; , \quad
\widetilde  \varepsilon_{\widetilde 1} := 
\frac{
\Gamma_{\widetilde \nu^1_{\mathrm{R}} \widetilde L} - \Gamma_{\widetilde
\nu^{1*}_{\mathrm{R}}
\widetilde{L}^*}
}{
\Gamma_{\widetilde \nu^1_{\mathrm{R}} \widetilde L} + \Gamma_{\widetilde
\nu^{1*}_{\mathrm{R}} 
\widetilde{L}^*}
}\; 
\end{eqnarray}
for the decay of $\nu^1_{\mathrm{R}}$ into slepton and
 Higgsino and for the decays of the sneutrino $\widetilde \nu^1_{\mathrm{R}}$. 
At tree level, the decay rates are
\begin{eqnarray}
\lefteqn{\Gamma_{\nu^1_{\mathrm{R}} L} + \Gamma_{\nu^1_{\mathrm{R}} \overline L}
=
\Gamma_{\nu^1_{\mathrm{R}} \widetilde L} + \Gamma_{\nu^1_{\mathrm{R}} 
\widetilde{L}^*}
\;=\;
\Gamma_{\widetilde \nu^{1*}_{\mathrm{R}} L}
\;=\;
\Gamma_{\widetilde \nu^1_{\mathrm{R}} \overline L}
\;=\;
\Gamma_{\widetilde \nu^1_{\mathrm{R}} \widetilde L}
\;=\;
\Gamma_{\widetilde \nu^{1*}_{\mathrm{R}} \widetilde{L}^*}
}\nonumber \\
&&=\;
\frac{M_{\mathrm{R}1}}{8 \pi}(Y^\dagger_\nu Y_\nu)_{11} 
\, .\! \!
\end{eqnarray}
The contributions to the decay asymmetries 
arise from the interference 
of the diagrams for the tree-level decays with the loop diagrams.
The one-loop diagrams for the decay 
$\nu^1_{\mathrm{R}}\rightarrow L^f_a H_\mathrm{u}{}_b$    
 are shown in figure \ref{fig:Leptogenesis_Triplet_SUSY}. 
Compared to the
supersymmetric type I see-saw case, there are additional diagrams contributing
to $\varepsilon^{\mathrm{MSSM}}_1$, 
 \ref{fig:Leptogenesis_Triplet_SUSY}(c) 
and \ref{fig:Leptogenesis_Triplet_SUSY}(f), which involve the triplet Higgs 
and its superpartner. 
The additional diagrams corresponding to the 
decay of $\nu^1_{\mathrm{R}}$ into slepton 
and Higgsino and to the decays of the sneutrino 
$\widetilde \nu^1_{\mathrm{R}}$ are not shown explicitly, 
but are included in the analysis.

Using FeynCalc \cite{Mertig:1991an}, the calculation of the decay asymmetries
corresponding to the diagrams \ref{fig:Leptogenesis_Triplet_SUSY}(c) and 
\ref{fig:Leptogenesis_Triplet_SUSY}(f) of 
figure \ref{fig:Leptogenesis_Triplet_SUSY} yields 
\begin{subequations}\label{eq:DecayAsTriplet}\begin{eqnarray}\label{eq:EffDecayAss_MinTypeIIScenarios}
\varepsilon^{(c)}_{1} &\!\!=\!\!& -\frac{3}{8 \pi} 
 \frac{M_{\mathrm{R}1}}{v^2_\mathrm{u}}
  \frac{\sum_{fg}\mbox{Im}\, [(Y^*_\nu)_{f1} (Y^*_\nu)_{g1} 
 (m_{\mathrm{LL}}^{\mathrm{II}})_{fg}]}{(Y_\nu^\dagger Y_\nu)_{11}}
  \, y \,\left[ - 1 + y \ln \left( \frac{y +1}{y}\right) \!\right]
 \!,\\
\varepsilon^{(f)}_{1} &\!\!=\!\!& -\frac{3}{8 \pi} 
 \frac{M_{\mathrm{R}1}}{v^2_\mathrm{u}}
  \frac{\sum_{fg}\mbox{Im}\, [(Y^*_\nu)_{f1} (Y^*_\nu)_{g1} 
 (m_{\mathrm{LL}}^{\mathrm{II}})_{fg}]}{(Y_\nu^\dagger Y_\nu)_{11}}
 \, y \,\left[1 - (1+y) \ln \left( \frac{y +1}{y}\right) \!\right]
 \!.
\end{eqnarray}\end{subequations}
We have defined $y := M^2_\Delta / M^2_{\mathrm{R}1}$. For dealing with 
lepton number violating interactions, we use the methods derived 
in \cite{Denner:1992vz}. 
The results for the contributions to the decay asymmetries from the triplet in 
the SM and from the triplet superfields in the MSSM are
\begin{subequations}\begin{eqnarray}
\varepsilon^{\mathrm{SM},\mathrm{II}}_{1}  &=& 
\varepsilon^{(c)}_{1}\; , \\
\varepsilon^{\mathrm{MSSM},\mathrm{II}}_{1}  &=& 
\varepsilon^{(c)}_{1} + \varepsilon^{(f)}_{1}
\; .
\end{eqnarray}\end{subequations}
The MSSM results are new. In the SM, we correct the previous 
result of \cite{Hambye:2003ka} by a factor of $-3/2$. 
As we will see below, our results in 
the limit $y \gg 1$ agree with the calculation in the  
effective approach, where the particles much heavier than $M_{\mathrm{R}1}$ are
integrated out.   
In the MSSM, we furthermore obtain 
\begin{eqnarray}
\varepsilon{}^{\mathrm{MSSM},\mathrm{II}}_{1} = 
\widetilde \varepsilon {}\,^{\mathrm{MSSM},\mathrm{II}}_{1} = 
\varepsilon{}^{\mathrm{MSSM},\mathrm{II}}_{\widetilde 1} = 
\widetilde \varepsilon {}\,^{\mathrm{MSSM},\mathrm{II}}_{\widetilde 1}\; .
\end{eqnarray}

In addition, we reproduce the known results \cite{Covi:1996wh}
for the decay asymmetries corresponding to the diagrams (a), (b), (d) and (e) 
which contribute to $\varepsilon_1^\mathrm{I}$ in the SM and in the MSSM:
\begin{subequations}\label{eq:EffDecayAss_MinTypeIIScenarios}\begin{eqnarray}
\varepsilon^{(a)}_{1} &=& \frac{1}{8 \pi} 
 \frac{\sum_{j\not=1} \mbox{Im}\, [(Y^\dagger_\nu Y_\nu)^2_{1j}]}{
 \sum_f \, |(Y_\nu)_{f1}|^2} 
 \, \sqrt{x_j} \,\left[1 - (1+x_j) \ln \left( \frac{x_j +1}{x_j}\right) \right]
 ,\\
 \varepsilon^{(b)}_{1} &=& \frac{1}{8 \pi} 
 \frac{\sum_{j\not=1} \mbox{Im}\, [(Y^\dagger_\nu Y_\nu)^2_{1j}]}{
 \sum_f \, |(Y_\nu)_{f1}|^2} 
 \, \sqrt{x_j} \,\left[  \frac{1}{1-x_j} \right]
 ,\\
 \varepsilon^{(d)}_{1} &=& \frac{1}{8 \pi} 
 \frac{\sum_{j\not=1}\mbox{Im}\, [(Y^\dagger_\nu Y_\nu)^2_{1j}]}{
 \sum_f \, |(Y_\nu)_{f1}|^2} 
 \, \sqrt{x_j} \,\left[ - 1 + x_j \ln \left( \frac{x_j +1}{x_j}\right) \right]
 ,\\
  \varepsilon^{(e)}_{1} &=& \frac{1}{8 \pi} 
 \frac{\sum_{j\not=1} \mbox{Im}\, [(Y^\dagger_\nu Y_\nu)^2_{1j}]}{
 \sum_f \, |(Y_\nu)_{f1}|^2} 
 \, \sqrt{x_j} \,\left[  \frac{1}{1-x_j} \right],
\end{eqnarray}\end{subequations}
with $x_j := M^2_{\mathrm{R}j} / M^2_{\mathrm{R}1}$ for $j \not= 1$. 
The results for the type I contribution to the decay asymmetries  
in the SM and in the MSSM are
\begin{subequations}\begin{eqnarray}
\varepsilon^{\mathrm{SM},\mathrm{I}}_{1}  &=& 
\varepsilon^{(a)}_{1} + \varepsilon^{(b)}_{1} \; , \\
\varepsilon^{\mathrm{MSSM},\mathrm{I}}_{1}  &=& 
\varepsilon^{(a)}_{1} + \varepsilon^{(b)}_{1} 
+ \varepsilon^{(d)}_{1} +
\varepsilon^{(e)}_{1}\;.
\end{eqnarray}\end{subequations}
In the MSSM, the remaining decay asymmetries are equal 
to $\varepsilon{}^{\mathrm{MSSM},\mathrm{I}}_{1}$ \cite{Covi:1996wh},  
\begin{eqnarray}
\varepsilon{}^{\mathrm{MSSM},\mathrm{I}}_{1} = 
\widetilde \varepsilon{}\,^{\mathrm{MSSM},\mathrm{I}}_{1} = 
\varepsilon{}^{\mathrm{MSSM},\mathrm{I}}_{\widetilde 1} = 
\widetilde \varepsilon{}\,^{\mathrm{MSSM},\mathrm{I}}_{\widetilde 1}\; .
\end{eqnarray}
Note that the type I results can be brought to a form which contains the 
neutrino mass matrix using  
\begin{eqnarray}
\frac{1}{8 \pi} 
 \frac{\sum_{j\not=1} \mbox{Im}\, [(Y^\dagger_\nu Y_\nu)^2_{1j}]}{
 \sum_f \, |(Y_\nu)_{f1}|^2} \, \frac{1}{\sqrt{x_j}} =
 -\frac{1}{8 \pi} 
 \frac{M_{\mathrm{R}1}}{v^2_\mathrm{u}}
  \frac{\sum_{fg}\mbox{Im}\, [(Y^*_\nu)_{f1} (Y^*_\nu)_{g1} 
 (m_{\mathrm{LL}}^{\mathrm{I}})_{fg}]}{(Y_\nu^\dagger Y_\nu)_{11}}\,.
\end{eqnarray}
In the limit $y \gg 1$ and $x_j \gg 1$ for all $j \not= 1$, 
which corresponds to a large gap between the mass  
$M_{\mathrm{R}1}$  and
the masses $M_{\mathrm{R}2}$, $M_{\mathrm{R}3}$ and $M_\Delta$, using 
\begin{subequations}\begin{eqnarray}
z \,\left[1 - (1+z) \ln \left( \frac{z +1}{z}\right) \right] 
&\stackrel{z \gg 1}{\rightarrow}& - \frac{1}{2} \; , \\
z \,\left[  \frac{1}{1-z} \right]
&\stackrel{z \gg 1}{\rightarrow}& - 1 \; , \\
 z \,\left[ - 1 + z\ln \left( \frac{z +1}{z}\right) \right]
 &\stackrel{z \gg 1}{\rightarrow}& - \frac{1}{2}   
\end{eqnarray}\end{subequations}
for $z\in\{y,x_j\}$, we obtain the simple results for the decay
asymmetries $\varepsilon^{\mathrm{SM}}_{1}  =  
\varepsilon^{\mathrm{SM,I}}_{1} +  \varepsilon^{\mathrm{SM,II}}_{1}$ 
and $\varepsilon^{\mathrm{MSSM}}_{1} =
\varepsilon^{\mathrm{MSSM,I}}_{1} +  \varepsilon^{\mathrm{MSSM,II}}_{1}$, 
\begin{subequations}\begin{eqnarray}
\varepsilon^{\mathrm{SM}}_{1}  &=& 
  \frac{3}{16 \pi} 
 \frac{M_{\mathrm{R}1}}{v_\mathrm{u}^2} 
 \frac{\sum_{fg}\mbox{Im}\, [(Y^*_\nu)_{f1} (Y^*_\nu)_{g1} 
(m^{\mathrm{I}}_{\mathrm{LL}}+ m^{\mathrm{II}}_{\mathrm{LL}})_{fg}]}{\sum_h 
\, |(Y_\nu)_{h1}|^2}\; ,\\
\varepsilon^{\mathrm{MSSM}}_{1} &=&
  \frac{3}{8 \pi} 
 \frac{M_{\mathrm{R}1}}{v_\mathrm{u}^2} 
 \frac{\sum_{fg}\mbox{Im}\, [(Y^*_\nu)_{f1} (Y^*_\nu)_{g1} 
(m^{\mathrm{I}}_{\mathrm{LL}}+ m^{\mathrm{II}}_{\mathrm{LL}})_{fg}]}{\sum_h \,
|(Y_\nu)_{h1}|^2}\;. 
\end{eqnarray}\end{subequations} 
In the presence of such a mass gap, the calculation  
can also be performed in an effective approach 
after integrating out the two heavy right-handed neutrinos and the heavy 
triplet, generating contributions to the effective neutrino mass operator, 
as we now discuss.

\section{Effective Approach to Type II Leptogenesis}\label{sec:EffLeptogenesis}

In the SM and the MSSM, viewed as effective theories, neutrino masses can be 
introduced via the lowest dimensional effective neutrino mass operator 
\label{eq:The3FormsOfNuMassOp}\begin{subequations}\begin{eqnarray}
\label{eq:The3FormsOfNuMassOp_SM} \mathscr{L}_{\kappa}^{\mathrm{SM}}  
 &=&\frac{1}{4} 
 \kappa_{gf} \, (\overline{L^\mathrm{C}}^g \cdot H)\, 
 \, (L^{f} \cdot H) 
  +\text{h.c.} \; , \\
\label{eq:The3FormsOfNuMassOp_MSSM}   \mathscr{L}_{\kappa}^{\mathrm{MSSM}} 
 &=&-\frac{1}{4} 
  {\kappa}^{}_{gf} \, (\SuperField{L}^{g}\cdot
 \SuperField{H}_\mathrm{u}) 
 \, (\SuperField{L}^{f}\cdot \SuperField{H}_\mathrm{u})\, \big|_{\theta\theta} 
 +\text{h.c.} \;.
\end{eqnarray}\end{subequations}
After electroweak symmetry breaking, the effective operator yields    
Majorana masses for the light neutrinos, 
\begin{equation}
\mathcal{L}_\nu\;=\;-\tfrac{1}{2} m_{LL}^\nu \overline \nu_{\mathrm{L}} 
\nu^{\ChargeC f}_{\mathrm{L}} \;,\;\;
\mbox{with}\;\; m{}_{\mathrm{LL}}^\nu \;=\; - \frac{v_\mathrm{u}^2}{2} (\kappa)^*\;.
\end{equation}
Let us assume that the lepton asymmetry is generated via the decay of the
lightest right-handed neutrino and that all other additional particles, in
particular the ones which generate the type II contribution, are much
heavier than $M_{\mathrm{R}1}$. 
We furthermore assume that we can neglect their   
population in the early universe, 
e.g.\ that their masses are much larger than the reheating temperature $T_R$ and
that they are not produced non-thermally in a large amount. We also assume that  
 they approximately do not contribute to washout processes. 
This scenario is motivated by supersymmetric GUTs, where 
additional charged particles like e.g.\ SU(2)$_{\mathrm{L}}$-triplets 
with intermediate scale masses could spoil gauge coupling
unification at $M_{\mathrm{GUT}} \approx 2\cdot 10^{16}$ GeV.

\begin{figure}
 \centering
 \includegraphics[scale=1,angle=0]{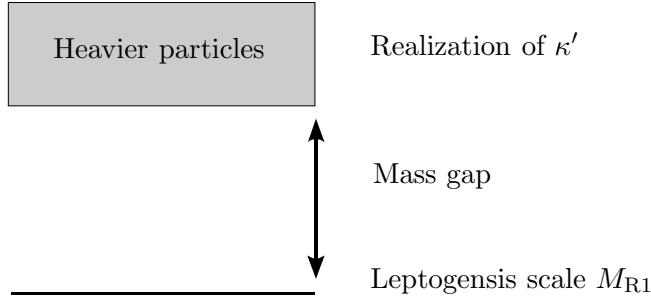}
 \caption{\label{fig:gap}
 Graphical illustration of the effective description of neutrino masses at 
 the leptogenesis scale.  
 }
\end{figure}

For a minimal effective approach, it is convenient to isolate the type I 
contribution 
from the lightest right-handed neutrino as follows: 
\begin{equation}
m^{\nu}_{\mathrm{LL}} \;=\; - \frac{v_\mathrm{u}^2}{2} \left[ 
 2 (Y_{\nu})_{f1} M_{\mathrm{R}1}^{-1} (Y^T_{\nu})_{1f} + 
 \kappa'^*\right] .
\end{equation} 
  $\kappa'$ includes type I contributions from the heavier
 right-handed neutrinos, plus any additional (type II) contributions from 
 heavier particles. Examples for realizations of the neutrino mass operator
 can be found e.g.\ in \cite{Ma:1998dn}.
At $M_{\mathrm{R}1}$, 
the most minimal extension of the SM or the MSSM
 would then be to introduce the effective neutrino mass operator $\kappa'$ plus 
one right-handed neutrino
$\nu_\mathrm{R}^1$ with mass $M_{\mathrm{R}1}$ and Yukawa couplings 
$(Y_{\nu})_{f1}$ to the lepton doublets $L^f$, defined as 
$(Y_\nu)_{f 1}(\SuperField{L}^{f}\cdot
 \SuperField{H}_\mathrm{u})\,  \SuperField{\nu}^{\chargec 1} 
$ in the superpotential of the MSSM and 
$
- (Y_\nu)_{f 1}({L}^{f}\cdot
 H)\,  \nu_\mathrm{\mathrm{R}}^{1} 
$ in Lagrangian of the SM. 
The situation is illustrated in figure \ref{fig:gap}.

\vspace{3mm}

\subsection{Decay Asymmetries for the SM and MSSM}
\begin{figure}[htb]
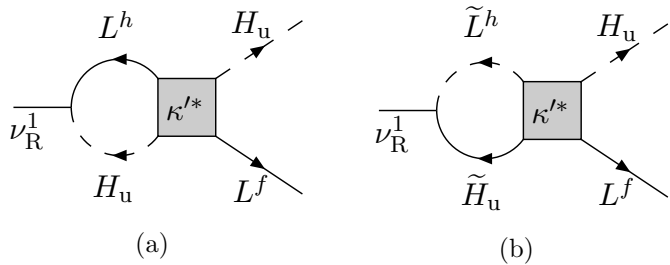

 \centering  
 \subfigure[]{$\CenterEps[1]{lgdiag_xSM}$}
 \quad\quad
 \subfigure[]{$\CenterEps[1]{lgdiag_xSUSY}$}
 \caption{\label{fig:EffLeptogenesis_SUSY}
 Loop diagrams contributing to the decay asymmetry via the decay  
 $\nu^1_{\mathrm{R}}\rightarrow L^f_a H_\mathrm{u}{}_b$ in the 
 MSSM with a (lightest) right-handed neutrino 
 $\nu^1_{\mathrm{R}}$ and a  
 neutrino mass matrix determined by $\kappa'$. 
 Further contributions to the generated baryon asymmetry stem from the decay
 of $\nu^1_{\mathrm{R}}$ into slepton and
 Higgsino and from the decays of the sneutrino $\widetilde \nu^1_{\mathrm{R}}$. 
 With $H_\mathrm{u}{}$ 
 renamed to the SM Higgs, the first diagram contributes in the extended SM. 
 }
\end{figure}

The contributions to the decay asymmetries in the effective approach 
stem from the interference 
of the diagram for the tree-level decay with the loop diagrams containing the
effective operator.
 In the SM, 
 the interference with diagram (a) of figure 
\ref{fig:EffLeptogenesis_SUSY} gives the simple result 
\begin{eqnarray}\label{eq:EffDecayAss_SM}
\varepsilon{}^{\mathrm{SM}}_{1} =
  \frac{3}{16 \pi} 
 \frac{M_{\mathrm{R}1}}{v_\mathrm{u}^2} 
 \frac{\sum_{fg}\mbox{Im}\, [(Y^*_\nu)_{f1} (Y^*_\nu)_{g1} 
 (m{}_{\mathrm{LL}}^\nu)_{fg}]}{(Y_\nu^\dagger Y_\nu)_{11}} 
  =:
 - \frac{3}{16 \pi} 
 \frac{M_{\mathrm{R}1}}{v_\mathrm{u}^2}  \<m^{\mathrm{BAU}}\>
 \;,
\end{eqnarray}
where we have
introduced the effective mass for leptogenesis $ \<m^{\mathrm{BAU}}\>$,  
\begin{eqnarray}\label{eq:EffMassForLeptogenesis}
\<m^{\mathrm{BAU}}\>:= - \frac{\sum_{fg}\mbox{Im}\, [(Y^*_\nu)_{f1} (Y^*_\nu)_{g1} 
 (m{}_{\mathrm{LL}}^\nu)_{fg}]}{(Y_\nu^\dagger Y_\nu)_{11}} \; .
\end{eqnarray}
For the
supersymmetric case, diagram (a) and diagram (b) contribute to 
$\varepsilon_{1}$ and we obtain: 
\begin{eqnarray}\label{eq:EffDecayAss_MSSM}
\varepsilon^{\mathrm{MSSM}}_{1} =
  \frac{3}{8 \pi} 
 \frac{M_{\mathrm{R}1}}{v_\mathrm{u}^2} 
 \frac{\sum_{fg}\mbox{Im}\, [(Y^*_\nu)_{f1} (Y^*_\nu)_{g1} 
 (m{}_{\mathrm{LL}}^\nu)_{fg}]}{(Y_\nu^\dagger Y_\nu)_{11}} 
 =:
 - \frac{3}{8 \pi} 
 \frac{M_{\mathrm{R}1}}{v_\mathrm{u}^2}  \<m^{\mathrm{BAU}}\>
 .
\end{eqnarray}
Explicit calculation furthermore yields 
\begin{eqnarray}\label{eq:SusyDecayAsymmetries}
\varepsilon{}^{\mathrm{MSSM}}_{1} =\widetilde \varepsilon{}\,^{\mathrm{MSSM}}_{1} =
\varepsilon{}^{\mathrm{MSSM}}_{\widetilde 1} =
\widetilde \varepsilon{}\,^{\mathrm{MSSM}}_{\widetilde 1}\;.
\end{eqnarray} 
The results are independent of the details of the realization of the neutrino
mass operator $\kappa'$.  
Note that, since the diagrams where the lightest right-handed neutrino runs in the loop 
do not contribute to leptogenesis, we have written  
$m_{\mathrm{LL}}^\nu = - \frac{v_\mathrm{u}^2}{2} (\kappa)^*$ instead of 
$m'{}_{\mathrm{LL}}^\nu:= - \frac{v_\mathrm{u}^2}{2} (\kappa')^*$ in the 
formulae (\ref{eq:EffDecayAss_SM}) 
- (\ref{eq:EffDecayAss_MSSM}). Having done this, the decay asymmetries are then
seen to be directly related to the neutrino mass matrix $m_{\mathrm{LL}}^\nu$.  

For neutrino masses via the type I see-saw mechanism, they are in agreement 
with the known results \cite{Covi:1996wh} 
(equation (\ref{eq:EffDecayAss_MinTypeIIScenarios})) in the limit 
$M_{\mathrm{R}2},M_{\mathrm{R}3} \gg M_{\mathrm{R}1}$. 
In the limit $M_\Delta \gg M_{\mathrm{R}1}$, the results obtained in the 
effective approach are also in agreement with our full theory calculation 
in the minimal type II scenarios with SU(2)$_{\mathrm{L}}$-triplets in equation 
(\ref{eq:DecayAsTriplet}). In
particular, we confirm the correction by the factor $-3/2$ 
compared to the previous result of \cite{Hambye:2003ka} in the SM.

\subsection{The Produced Baryon Asymmetry}
 The generated B-L asymmetry, i.e.~the ratio of the number density over the
 entropy density $Y_{\mathrm{B-L}}=n_{\mathrm{B-L}}/s$, 
 can be written as 
 \begin{subequations}\label{eq:LeptAs}\begin{eqnarray}
 \label{eq:LeptAs_SM}Y_\mathrm{B-L}^{\mathrm{SM}} &=& -\eta\,
 \varepsilon_{1} \, Y^{\mathrm{eq}}_{\nu^1_{\mathrm{R}}}  \; ,\\
 \label{eq:LeptAs_MSSM}Y_\mathrm{B-L}^{\mathrm{MSSM}} &=& 
 -\eta \left[ 
 \tfrac{1}{2}(\varepsilon_{1}+\widetilde \varepsilon_{1}) \, Y^{\mathrm{eq}}_{\nu^1_{\mathrm{R}}}
 + 
 \tfrac{1}{2}(\varepsilon_{\widetilde 1} + \widetilde \varepsilon_{\widetilde 1})\, Y^{\mathrm{eq}}_{\widetilde \nu^1_{\mathrm{R}}}
  \right] .
 \end{eqnarray}\end{subequations}
 $\varepsilon_{1}$  (and $\widetilde \varepsilon_{ 1}$) are the decay 
 asymmetries of the lightest right-handed
 neutrino into (s)lepton and Higgs(ino) and 
 $\varepsilon_{\widetilde 1}$ (and $\widetilde \varepsilon_{\widetilde 1}$) are
 the decay asymmetries of the lightest right-handed
 sneutrino. Ignoring supersymmetry breaking, 
 the right-handed neutrinos and sneutrinos
 have equal mass $M_{\mathrm{R}1}$.  
 $Y^{\mathrm{eq}}_{\nu^1_{\mathrm{R}}}$ and 
 $Y^{\mathrm{eq}}_{\widetilde \nu^1_{\mathrm{R}}}$ are the 
 number densities of the neutrino and sneutrino at $T \gg M_{\mathrm{R}1}$ 
 if they were in thermal 
 equilibrium, normalized with respect to the entropy
 density. They are given by
\begin{eqnarray}
 Y^{\mathrm{eq}}_{\nu^1_{\mathrm{R}}} \approx \frac{45 \,\zeta (3)}{ \pi^4
 g_* k}\frac{3}{4} 
  \;\; \mbox{and} \;\; 
 Y^{\mathrm{eq}}_{\widetilde \nu^1_{\mathrm{R}}} \approx \frac{45 \,\zeta (3)}{ \pi^4
 g_* k}
 \; ,
 \end{eqnarray}
where $g^*$ is the effective number of 
 degrees of freedom, which amounts 
 $106.75$ in the SM and $228.75$ in the MSSM, and $k$ is the Boltzmann constant.
 
 Equation (\ref{eq:LeptAs}) also provides the definition for the efficiency
 factor $\eta$ for leptogenesis. It
 can be computed from a set of coupled Boltzmann equations
 (see e.g.\  \cite{Buchmuller:2000as}) and 
 it is subject to e.g.\ thermal
 correction \cite{Giudice:2003jh} and corrections from spectator processes 
 \cite{Buchmuller:2001sr}, 
 $\Delta\mathrm{L} \!=\! 1$ processes involving gauge bosons 
 \cite{Pilaftsis:2003gt,Giudice:2003jh} and from renormalization group
 running \cite{Barbieri:1999ma,Antusch:2003kp}.  
 In the effective approach and for thermal leptogenesis with a reheating
 temperature $T_\mathrm{R} \gg M_{\mathrm{R}1}$, which is most
 independent of the cosmological model and of the model for neutrino masses, 
 we assume that to a good approximation  
 the efficiency factor depends only on the quantity $\widetilde m_1$ 
 \cite{Buchmuller:2000as}, defined by 
\begin{eqnarray}
\widetilde m_1 := \frac{\sum_{f} (Y^\dagger_\nu)_{1f} (Y_\nu)_{f1}\, 
v_\mathrm{u}^2 }{ M_{\mathrm{R1}} } \; , 
\end{eqnarray} 
and on the initial population of right-handed (s)neutrinos.  
This means, we neglect e.g.\ the contribution to washout processes from 
diagrams involving the additional particles which are involved in realizing the
effective operator $\kappa'$. 
For example, in type I see-saw scenarios  
the effects from the heavier right-handed neutrinos and their
Yukawa couplings can be neglected if $M_{\mathrm{R}1}$ is 
much smaller than $10^{14}$ GeV. Under this assumption, we can use the 
results for $\eta$ from type I see-saw models. 
See e.g.\ \cite{Giudice:2003jh} for figures showing $\eta (\widetilde m_1)$ for various 
initial populations of right-handed (s)neutrinos. 
A population of right-handed (s)neutrinos required for 
leptogenesis can also be 
produced non-thermally, e.g.\ via the decay of the inflaton. Such scenarios 
depend on the specific cosmological model. They 
could be very efficient, since $\nu^1_{\mathrm{R}}$ and 
$\widetilde \nu^1_{\mathrm{R}}$ would be almost completely out-of-equilibrium
when they decay.

 From the decay of the right-handed (s)neutrinos, a lepton asymmetry 
 is produced which is then
 transformed into a baryon asymmetry via sphaleron transitions. Since the 
 latter conserve B-L, we write the negative of the lepton asymmetry as B-L 
 asymmetry in equation (\ref{eq:LeptAs}).   
  The baryon asymmetry is then related to the B-L asymmetry  
  $Y_\mathrm{B}$ via 
 \begin{eqnarray}
 Y_\mathrm{B} = \alpha \, Y_\mathrm{B-L}\; , \;\; 
 \mbox{with}\;\; \alpha \approx \frac{24 + 4 N_H}{66 + 13 N_H}
 \end{eqnarray}   
 and with $N_H$ being the number of Higgs doublets. 
 With $\varepsilon_{1}=\widetilde \varepsilon_{1}=
 \varepsilon_{\widetilde 1}=\widetilde \varepsilon_{\widetilde 1}$ 
 from equation
 (\ref{eq:SusyDecayAsymmetries}) in the MSSM 
 and using $s/n_\gamma\approx 7.04 k$, 
 the produced baryon asymmetry in terms of the baryon to photon ratio 
 in the SM and in the MSSM is 
 approximately given by 
 \begin{subequations}\label{eq:BAU_epsilon_eta}\begin{eqnarray}
  \frac{n_\mathrm{B}^{\mathrm{SM}}}{n_\gamma}
 &\approx& - \,0.97\cdot10^{-2}\,\varepsilon_{1}\, 
 \eta\; ,\\
  \frac{n_\mathrm{B}^{\mathrm{MSSM}}}{n_\gamma}
 &\approx& - \,1.04\cdot10^{-2}\,\varepsilon_{1}
 \, \eta\; .
 \end{eqnarray}\end{subequations}
Note that the sign of the produced asymmetry is a relevant quantity here.  
$n_\mathrm{B}$ has to be positive since we calculate in the convention that we consist of
matter and not of anti-matter. In terms of the effective mass for leptogenesis, 
defined in equation (\ref{eq:EffMassForLeptogenesis}), 
$\<m^{\mathrm{BAU}}\> > 0$ is required for obtaining $n_\mathrm{B} > 0$.

\section{Type II Bound on Decay Asymmetry and on
$\boldsymbol{M_{\mathrm{R}1}}$}\label{sec:BoundonMR1}
In the effective approach, we can calculate a model-independent upper bound 
for the decay asymmetries $\varepsilon^\mathrm{SM}_{1}$ and 
$\varepsilon^\mathrm{MSSM}_{1}$ from the requirement of successful thermal
leptogenesis. For obtaining this bound, it is
useful to choose a basis where $m^\nu_{\mathrm{LL}}$ (and 
$M_{\mathrm{RR}}$) are diagonal. Note that the decay asymmetry is independent of
the basis for $Y_e$ . 
In this basis, we can write
\begin{eqnarray}
(Y_\nu)_{1f} = 
\begin{pmatrix}
\widetilde y_{11}\,e^{\I \phi_1} \\
\widetilde y_{21}\,e^{\I \phi_2}  \\
\widetilde y_{31}\,e^{\I \phi_3}  
\end{pmatrix}
, \quad
m^\nu_{\mathrm{LL}} = 
\begin{pmatrix}
m_1 & 0 & 0 \\
0 & m_2 &0 \\
0 & 0 & m_3 
\end{pmatrix},
\end{eqnarray} 
with real and positive $\widetilde y_{11}, \widetilde y_{12}$ and 
$\widetilde y_{13}$. For the effective mass for 
leptogenesis $\<m^{\mathrm{BAU}}\>$, defined in equation (\ref{eq:EffMassForLeptogenesis}), 
we obtain
 \begin{eqnarray}
\<m^{\mathrm{BAU}}\>&=& 
- \frac{\sum_{fg}\mbox{Im}\, [(Y^*_\nu)_{f1} (Y^*_\nu)_{g1} 
 (m{}_{\mathrm{LL}}^\nu)_{fg}]}{(Y_\nu^\dagger Y_\nu)_{11}} 
 \le 
 \frac{\widetilde y^2_{11} m_1 + \widetilde y^2_{21} m_2 + 
 \widetilde y^2_{31} m_3}{
 \widetilde y^2_{11} +  \widetilde y^2_{21} + \widetilde y^2_{31}} \nonumber \\
 &\le& m^\nu_\mathrm{max} 
 \; ,
\end{eqnarray}
with $m^\nu_\mathrm{max} :=\mbox{max}\,(m_1,m_2,m_3)$ being the 
largest neutrino mass at the energy scale $M_{\mathrm{R}1}$. 
Using equation (\ref{eq:EffDecayAss_SM}) and equation
(\ref{eq:EffDecayAss_MSSM}), this leads to the upper bounds
\begin{subequations}\label{eq:BoundsEffDecayAss}\begin{eqnarray}\label{eq:BoundEffDecayAss_SM}
|\varepsilon^{\mathrm{SM}}_{1}| 
&\le&
  \frac{3}{16 \pi} 
 \frac{M_{\mathrm{R}1}}{v_\mathrm{u}^2}  m^\nu_\mathrm{max}\; ,
 \\
\label{eq:BoundEffDecayAss_MSSM} |\varepsilon^{\mathrm{MSSM}}_{1}| 
&\le&
  \frac{3}{8 \pi} 
 \frac{M_{\mathrm{R}1}}{v_\mathrm{u}^2}  m^\nu_\mathrm{max} 
\end{eqnarray}\end{subequations}
for the decay asymmetries. 
Thus, the upper bound increases with increasing 
mass scale of the light neutrinos. 
Note that compared to the low energy value, 
the neutrino masses at the scale $M_{\mathrm{R}1}$ are enlarged by
renormalization group (RG) effects by   
$\approx +20 \% 
$ in the MSSM and $\approx +30 \%
$ in the SM, which raises the bounds on the decay asymmetries by the same values. 
More accurate results can be found e.g.\ in 
figure 4 of \cite{Antusch:2003kp}.

\begin{figure}[thb]
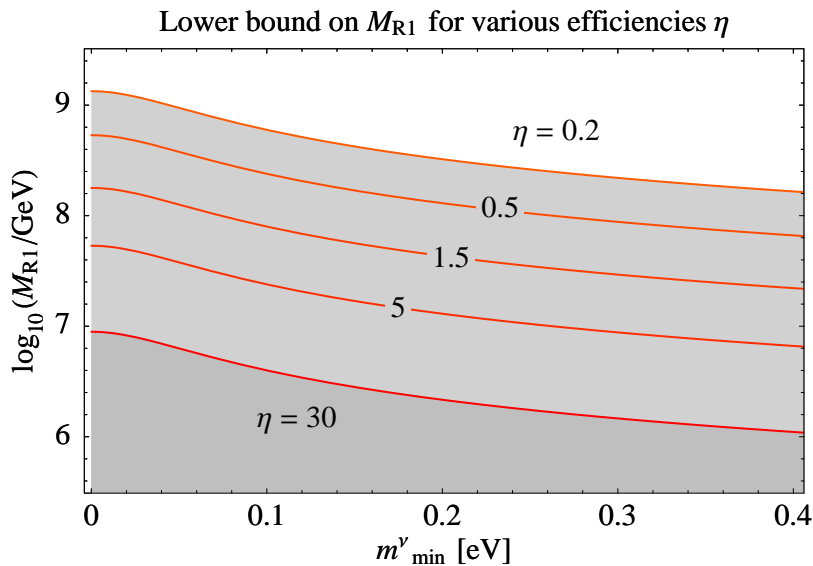

\centering
\CenterEps[0.95]{LowerBoundsMR1_TypeII}
 \caption{\label{fig:LowerBoundsMR1}
Graphical illustration of the lower bound on $M_{\mathrm{R}1}$ 
in the MSSM as a function of the mass of the lightest neutrino 
$m^\nu_\mathrm{min} := \mbox{min}\,(m_1,m_2,m_3)$ for some values of 
the efficiency factor $\eta$ and for a baryon to photon ratio  
$n_B = 6.5\cdot 10^{-10}$.  
In the extreme cases for thermal leptogenesis, the maximal value of the efficiency factor is 
$\eta^\mathrm{zero}_{\mathrm{max}} \approx 0.2$ for a zero initial 
population of 
$\nu_{\mathrm{R}1}$ and $\eta^\mathrm{dom.}_{\mathrm{max}}\approx
30$ for a maximal initial population (approximate values taken from \cite{Giudice:2003jh}).
 RG corrections to 
the neutrino masses at the scale $M_{\mathrm{R}1}$ of  
$\approx +20 \% 
$ in the MSSM and $\approx +30 \%
$ in the SM are included (see e.g.\ figure 4 in \cite{Antusch:2003kp}).    
 }
\end{figure}

Using equation (\ref{eq:BAU_epsilon_eta}), 
for a given efficiency factor $\eta$ and using an upper bound for 
$m^\nu_\mathrm{max}$, 
it can be transformed into a lower type II bound 
for the mass of the lightest right-handed neutrino: 
\begin{subequations}\label{eq:BoundsMR1}\begin{eqnarray}\label{eq:BoundEffDecayAss_SM}
M^{\mathrm{SM}}_{\mathrm{R}1}
&\ge&
  \frac{16 \pi}{3} 
 \frac{v_\mathrm{u}^2}{  m^\nu_\mathrm{max}} \frac{n_\mathrm{B}/n_\gamma }{0.97 \cdot
 10^{-2}\,\eta}\; ,
 \\
\label{eq:BoundEffDecayAss_MSSM} 
M^{\mathrm{MSSM}}_{\mathrm{R}1}
&\ge&
  \frac{8 \pi}{3} 
 \frac{v_\mathrm{u}^2}{  m^\nu_\mathrm{max}} \frac{n_\mathrm{B}/n_\gamma }{1.04 \cdot
 10^{-2}\,\eta}
 \; .
\end{eqnarray}\end{subequations}
The bound on $M_{\mathrm{R}1}$ is lower for a larger neutrino 
mass scale.  
It is shown in figure \ref{fig:LowerBoundsMR1} as a function of the 
neutrino mass scale, i.e.\ of the mass of
the lightest neutrino.  

The situation in the type II framework is very different to the type 
I see-saw case: 
E.g., for a normal mass ordering, 
the type II bound on the decay asymmetry is proportional to $m_3$,  
whereas the type I bound is
proportional to 
$\Delta m^2_{31}/m_3$. 
In addition, thermal type I leptogenesis gets less efficient for a larger 
neutrino mass scale since 
$\widetilde m_1 \ge m^\nu_\mathrm{min}$, with $m^\nu_\mathrm{min} :=
\mbox{min}\,(m_1,m_2,m_3)$. Together with an improved bound 
\cite{Buchmuller:2003gz} 
on the type I decay asymmetry, this strongly increases the type I bound on 
$M_{\mathrm{R}1}$ \cite{Davidson:2002qv} for 
increasing $m^\nu_\mathrm{min}$ and
finally leads to an upper bound for the absolute mass scale of the 
light neutrinos of $m^\nu_\mathrm{min} \le 0.12$ eV \cite{Buchmuller:2004nz}. 
In the type II scenario where $\widetilde m_1$ is in general independent of 
$m^\nu_\mathrm{max}$, there is no bound on the 
neutrino mass scale from the requirement of successful leptogenesis. 
A neutrino mass scale $\lesssim 0.35$ eV on the contrary allows for 
a mass of the lightest right-handed neutrino of about an order of magnitude 
below the bound in type I models, which might help thermal leptogenesis 
with respect to the gravitino problem  
in supersymmetric models. Note that in non-thermal leptogenesis scenarios, 
a lower bound on $M_{\mathrm{R}1}$ does in general 
not lead to a conflict with respect to the gravitino problem.

\section{Summary and Conclusions}
In this work, we have investigated type II leptogenesis 
via the decay of the lightest (s)neutrinos. 
In the MSSM with the type II contribution realized via an additional 
SU(2)$_\mathrm{L}$-triplet superfield, we have calculated the decay asymmetries    
for the lightest right-handed neutrino $\nu^1_{\mathrm{R}}$ and its 
superpartner $\widetilde \nu^1_{\mathrm{R}}$. 
In the SM, we have recalculated the decay asymmetry 
$\varepsilon^\mathrm{II}_1$   
and corrected the previous result. 
We have developed an effective
approach, assuming a gap between the mass $M_{\mathrm{R}1}$ of the lightest
(s)neutrino and the masses of the remaining particles involved in generating the
neutrino masses. 
We have calculated the effective decay asymmetries in the SM and in the MSSM. 
Leptogenesis in this framework is 
independent of the specific realization of the neutrino mass operator. The
total decay asymmetry $\varepsilon_1$ is proportional to the complete  
neutrino mass matrix
$m^{\nu}_{\mathrm{LL}}=m^{\mathrm{I}}_{\mathrm{LL}}+
m^{\mathrm{II}}_{\mathrm{LL}}$. 
We have derived a general upper bound 
(equation (\ref{eq:BoundsEffDecayAss})) on the total 
decay asymmetry and found that it increases with the neutrino 
mass scale, in sharp contrast to the type I case which leads to an 
upper bound of about 0.1 eV on the neutrino mass scale.  
It leads to a lower bound (equation (\ref{eq:BoundsMR1}) and figure 
\ref{fig:LowerBoundsMR1}) on the mass of the lightest right-handed neutrino 
significantly below the type I bound for partially degenerate neutrinos. 
Increasing the neutrino mass scale allows a lower reheat temperature, 
making thermal type II leptogenesis more consistent with the gravitino constraints 
in supersymmetric models.

\section*{Acknowledgements}
We would like to thank Michael Ratz for useful discussions.  
We acknowledge support from the PPARC grant PPA/G/O/2002/00468.

\providecommand{\bysame}{\leavevmode\hbox to3em{\hrulefill}\thinspace}

\end{document}